\documentclass[10pt,twocolumn,letterpaper]{article}

\usepackage{cvpr}
\usepackage{times}
\usepackage{epsfig}
\usepackage{graphicx}
\usepackage{amsmath}
\usepackage{amssymb}

\usepackage[breaklinks=true,bookmarks=false]{hyperref}

\cvprfinalcopy % *** Uncomment this line for the final submission

\def\cvprPaperID{****} % *** Enter the CVPR Paper ID here

\begin{document}

%%%%%%%%% TITLE
\title{MultiHead MultiModal Deep Interest Recommendation Network}

\author{Mingbao Yang, ShaoBo Li, Peng Zhou, Ansi Zhang, Yuanmeng Zhang\\
Guizhou University\\
Huaxi District, Guizhou University, Huaxi, Guiyang, Guizhou, China\\
{\tt\small dengjunchuang@gmail.com, lishaobo@gzu.edu.cn, pzhou@gzu.edu.cn}\\
{\tt\small zhangas@gzu.edu.cn, gs.ymzhang19@gzu.edu.cn}
%\and
%ShaoBo Li\\
%Guizhou University\\
%Huaxi District, Guizhou University, Huaxi, Guiyang, Guizhou, China\\
%{\tt\small lishaobo@gzu.edu.cn}
}

\maketitle
%\thispagestyle{empty}

%%%%%%%%% ABSTRACT
\begin{abstract}
   With the development of information technology, human beings are constantly producing a large amount of information at all times. How to obtain the information that users are interested in from the large amount of information has become an issue of great concern to users and even business managers. In order to solve this problem, from traditional machine learning to deep learning recommendation systems, researchers continue to improve optimization models and explore solutions. Because researchers have optimized more on the recommendation model network structure, they have less research on enriching recommendation model features, and there is still room for in-depth recommendation model optimization. Based on the DIN\cite{Authors01} model, this paper adds multi-head and multi-modal modules, which enriches the feature sets that the model can use, and at the same time strengthens the cross-combination and fitting capabilities of the model. Experiments show that the multi-head multi-modal DIN improves the recommendation prediction effect, and outperforms current state-of-the-art methods on various comprehensive indicators.
\end{abstract}

%%%%%%%%% BODY TEXT
\section{Introduction}

With the development of computer science and technology, people are producing a large amount of information at all times. Human beings have entered the era of information explosion\cite{Authors02,Authors03}. How to let users obtain the information they are interested in from the massive amount of information and how to improve the user's platform on the platform Resident time, how to improve users' product click-through rate, and conversion rate have become issues of close attention to major platforms and applications\cite{Authors04}.

In order to solve these problems, many companies, institutions, and research scholars have successively proposed various methods. The collaborative filtering algorithm\cite{Authors05,Authors06,Authors07} is undoubtedly the most successful one ever, but it is limited to most scenes and very sparse, and the prediction effect of the collaborative filtering algorithm is not very ideal. The matrix factorization algorithm\cite{Authors08} proposed subsequently solves this problem, and there are a large number of methods to solve the cold start problem\cite{Authors09,Authors10,Authors11,Authors12}, such as knowledge-based recommendation\cite{Authors13}, recommendation based on association rules\cite{Authors14}, recommendation based on content\cite{Authors15,Authors16}, recommendation based on knowledge graph\cite{Authors10,Authors17,Authors18}, and so on. However, the matrix factorization algorithm\cite{Authors08} is limited to its simple eigen-decomposition and interoperability, making it more and more difficult to meet the new era of information explosion. 

With the development of deep learning, researchers put forward the NeuralCF\cite{Authors19,Authors20} model, which divides the feature extraction module into user towers and item towers, and at the same time proposes Embedding layer to compress the original sparse onehot vectors, which greatly improves the efficiency of training and prediction, and at the same time interoperability the layer can be developed from a simple dot product operation to a complex MLP structure. NeuralCF stores the trained user and item characteristics in memory databases such as redis, which can be easily inferred online, but because the characteristics of the scene type are dynamic, it cannot be added to the user tower and the item tower. At this point, NeuralCF is slightly insufficient, while Embedding MLP\cite{Authors02} does not have this problem. In order to improve the prediction effect of the Embedding MLP model, researchers have proposed DeepFM\cite{Authors21}, xDeepFM\cite{Authors22}, Deep \& Cross Network (DCN)\cite{Authors23}, etc., so that features can be more effectively cross-combined based on human prior knowledge. At the same time, researchers found that different people have different degrees of interest in different products, and over time, interest will shift, so they proposed DIN\cite{Authors01} and DIEN\cite{Authors24} recommendation models. At the same time, with the development of reinforcement learning, researchers have also proposed Deep Reinforcement Network(DRN)\cite{Authors25,Authors26}. However, researchers have done less research on enriching the features of recommendation models, and there is still room for deep MLP (Multilayer Perception) recommendation model optimization.

Based on the observations above, this paper proposes a multi-head\cite{Authors28} multi-modal\cite{Authors04} deep interest recommendation network (MMDIN) to enhance the model's ability to extract features from data and enhance the ability of feature cross-combination; at the same time, it furtherly enriches the feature set that the model can use, so that the model prediction has more robust data features for supporting. This paper uses the MovieLens dataset to verify the effectiveness of the model. The main contributions of the MMDN model are as follows:

• On the basis of the DIN model, we added the MultiModal module to enable the model to utilize richer features and improve the prediction effect of the model.

• In the cross combination of features, we changed the MLP part to the ResNet\cite{Authors27} module, and added the Multi-Head\cite{Authors28} mechanism, so that as the number of layers increases, the model effect will not become worse, and at the same time the ability to extract features from multiple dimensions enhances the ability of feature cross combination and model fitting expression.

• The experiment did not choose low-performance python, disk-based mapreduce\cite{Authors29}, hive\cite{Authors30}, etc., but chose spark\cite{Authors31}, which is good at distributed memory computing, for data preprocessing, which improves the performance of data preprocessing.

%-------------------------------------------------------------------------
\section{MMDIN}
The overall structure of the MMDIN model is shown below in figure 1. The model mainly includes three key modules. They are the MultiModal module (the red dashed box in the figure 1), the Attention\cite{Authors32,Authors33} module (the yellow dashed box in the figure 1) and Multi-Head\cite{Authors28} ResNet\cite{Authors27} module (the part of the purple dashed box in the figure 1). In addition, the Attention module is basically the same as the Attention module of the DIN model. In the next paragraph, each module will be explained in detail.

\begin{figure}[h]
	\begin{center}
		\includegraphics[width=1.0\linewidth]{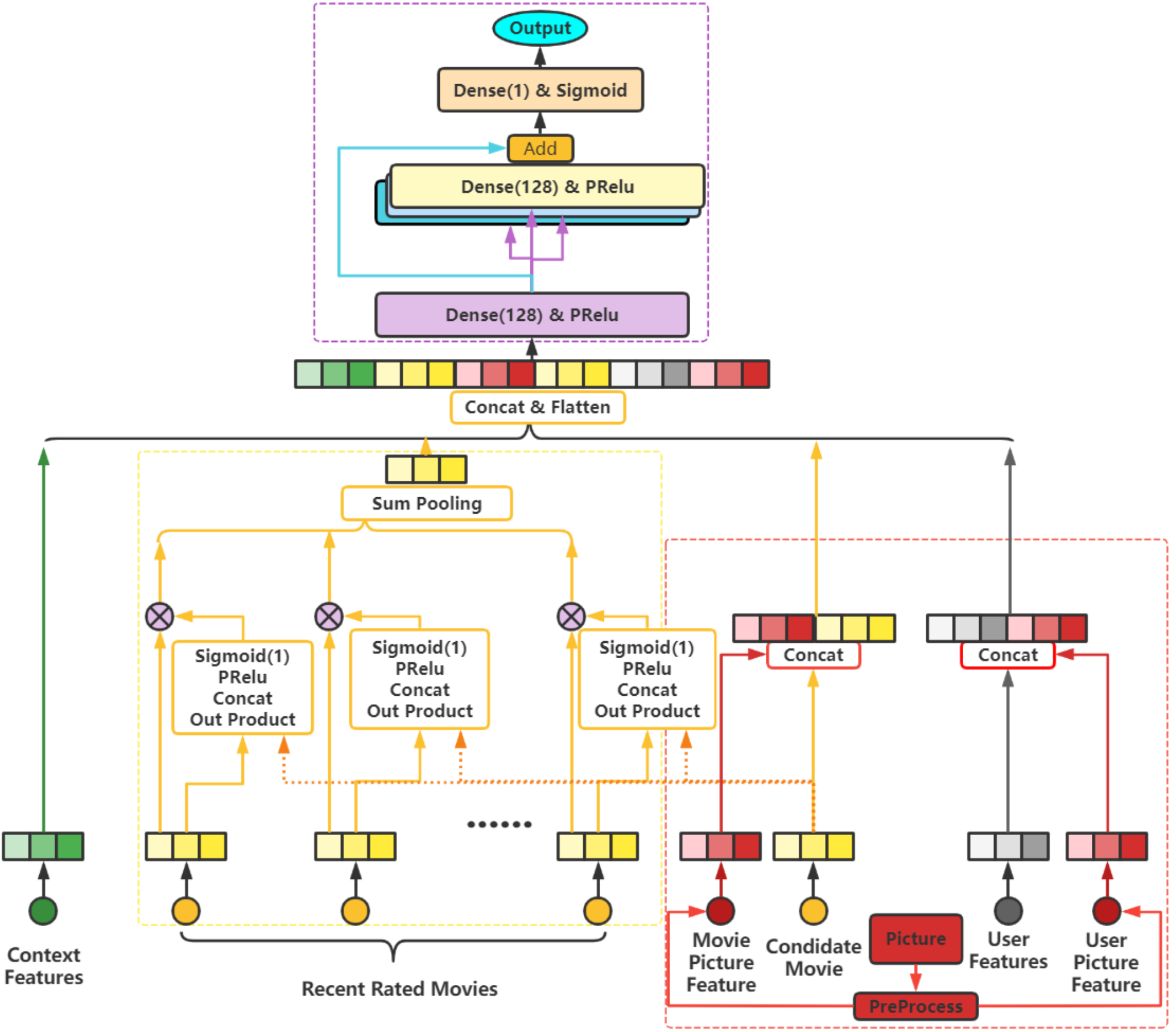}
	\end{center}
	\caption{Structure of MMDIN Model}
	\label{fig:long}
	\label{fig:onecol}
\end{figure}

%-------------------------------------------------------------------------
\subsection{MultiModal Module}
The MultiModal module is shown in the red dashed box in figure 1. It is mainly responsible for extracting the color characteristics of the posters, such as extracting the mean and standard deviation characteristics of the saturation, brightness, chroma, spatial frequency, RGB value and so on of the movie poster picture. It is more common that different age groups have different degrees of preference for color characteristics. For example, children tend to prefer colorful animations, young and middle-aged people prefer real life dramas, and older people prefer aging movies. Among them, the spatial frequency (SF) is an index that characterizes the clarity of the image. The larger the spatial frequency value, the clearer the image. The calculation formula is as follows:

\begin{equation}
SF = \sqrt {R{F^2} + C{F^2}}
\end{equation}

Among them, RF represents the row frequency of the image, and CF represents the column frequency of the image. The calculation formulas of them are as follows:

\begin{equation}
RF = \sqrt {\frac{1}{{IJ}}\sum\limits_{i = 1}^{I - 1} {\sum\limits_{j = 1}^{J - 1} {{{(F(i,j) - F(i,j + 1))}^2}} } }
\end{equation}

\begin{equation}
CF = \sqrt {\frac{1}{{IJ}}\sum\limits_{i = 1}^{I - 1} {\sum\limits_{j = 1}^{J - 1} {{{(F(i,j) - F(i + 1,j))}^2}} } }
\end{equation}

Among them, I and J represent the width and height pixel size of image F respectively, and F(i, j) represents the pixel value of image F in row i and column j.

In the model without the MultiModal module, the model can only use the characteristics of the movie's release year, user, movie and its rating, and there are fewer features that can be used; the MultiModal module obtains image saturation, brightness, chroma, and spatial frequency, RGB value and other features after preprocessing, and then statistical analysis to obtain the image features of the user’s historical rated movie, and then do one-hot and embedding, and connect to the movie features and user features respectively. The MultiModal Module enriches the feature set that can be used by the model, making the model has stronger expressive ability.

%-------------------------------------------------------------------------
\subsection{Attention Module}
The Attention module is shown in the yellow dashed box in figure 1, which is basically the same as the DIN model. The input is mainly divided into two parts: one part is movies recently rated by the user, and the other part is the candidate movie. In order to input ID-type features into the neural network for training, they need to be one-hot encoded firstly, and then connects to the embedding layer to convert it into a dense vector (as shown by the three side-by-side light yellow squares in figure 1) ; Then, the user history praise movie embeddings and candidate movie embedding are connected with the outer product\cite{Authors20} calculation results, and the attention weight is calculated through a multilayer neural network, and the original user history praise movie is weighted, and finally all user history praise movies are summed with weight and pooled. This is the structure of the Attention module. 

In the model without the attention module,the user's historical movie ratings are directly inputted into the summation pooling layer without discrimination, and it does not meet the characteristics of human attention with weight. In this model, the historical rating movie embeddings and the candidate movie embedding are connected to the outer product, and the relationship between the user's historical rating movies and the candidate movie are used as the weight, which makes the embeddings obtained by the final sum pooling more effective.

%-------------------------------------------------------------------------
\subsection{Multi-Head ResNet Module}
The Multi-Head ResNet module is shown in the purple dotted box in figure 1. The user characteristics, candidate movie characteristics, weighted sum pooled user history rating characteristics and scene characteristics are connected and flattened, and then input to Multi-Head ResNet module to make the final movie rating prediction.

The Multi-Head ResNet module uses dense function to compress the feature dimension to the specified dimension firstly, and then activates it with PRelu\cite{Authors34} and inputs it into ResNet. The ResNet structure adds the original features and the features processed by the hidden layer, so that the feature extracted by the multiple hidden layers will not become worse. The Multi-Head mechanism allows features to pass through multiple hidden layers to extract features of different dimensions at the same time, so that the model gets a better effect. After the features obtained above are added, they are connected to the output layer, and sigmoid activation is performed to obtain the final output rating result.

%------------------------------------------------------------------------
\section{Experiments}
%-------------------------------------------------------------------------
\subsection{Dataset Description}

The experiment evaluates the ratings of the first 1000 movies in the open source movie recommendation dataset MovieLens-265M, collects and analyzes the cover poster pictures of the first 1000 movies and inputs the MMDIN model for training and prediction. 

The dataset mainly includes movie dataset and rating dataset. The movie dataset mainly includes movie ID, movie titile, movie generes, movie release year, etc; the rating dataset mainly includes userID, movieID, movie rating, rating time, etc; the linking dataset mainly includes the movieID and its page link, which can be used to crawl its movie poster pictures.

With the aim of improving the performance, spark, which is good at distributed memory computing and tensorflow, which is good at building complex models, are used to clean the experimental dataset, preprocess the features and build the model. 

In terms of feature preprocessing, the experiment preprocesses and analyzes the movie rating times, film average rating, its standard deviation, film generes label, five films recently praised by users, user rating times, user average rating, average year of watched movies, standard deviation of year of watched movies, and the generes of five films recently praised by users.

Considering that users of different ages may have different preferences for image color, this experiment adds film cover image features, including saturation brightness, chroma, spatial frequency, mean and standard deviation of RGB, extracts and analyzes them, and finally inputs them into the MMDIN model for training and prediction.

A total of  5,048,339  rating data were divided into training dataset and test dataset in the proportion of 8:2 for training and model evaluation. Amony them, the distribution of positive and negative samples of training dataset and test dataset is as follows in figure 2. It can be seen that the proportion of positive and negative samples is relatively balanced.

\begin{figure}[h]
	\begin{center}
		\includegraphics[width=1.0\linewidth]{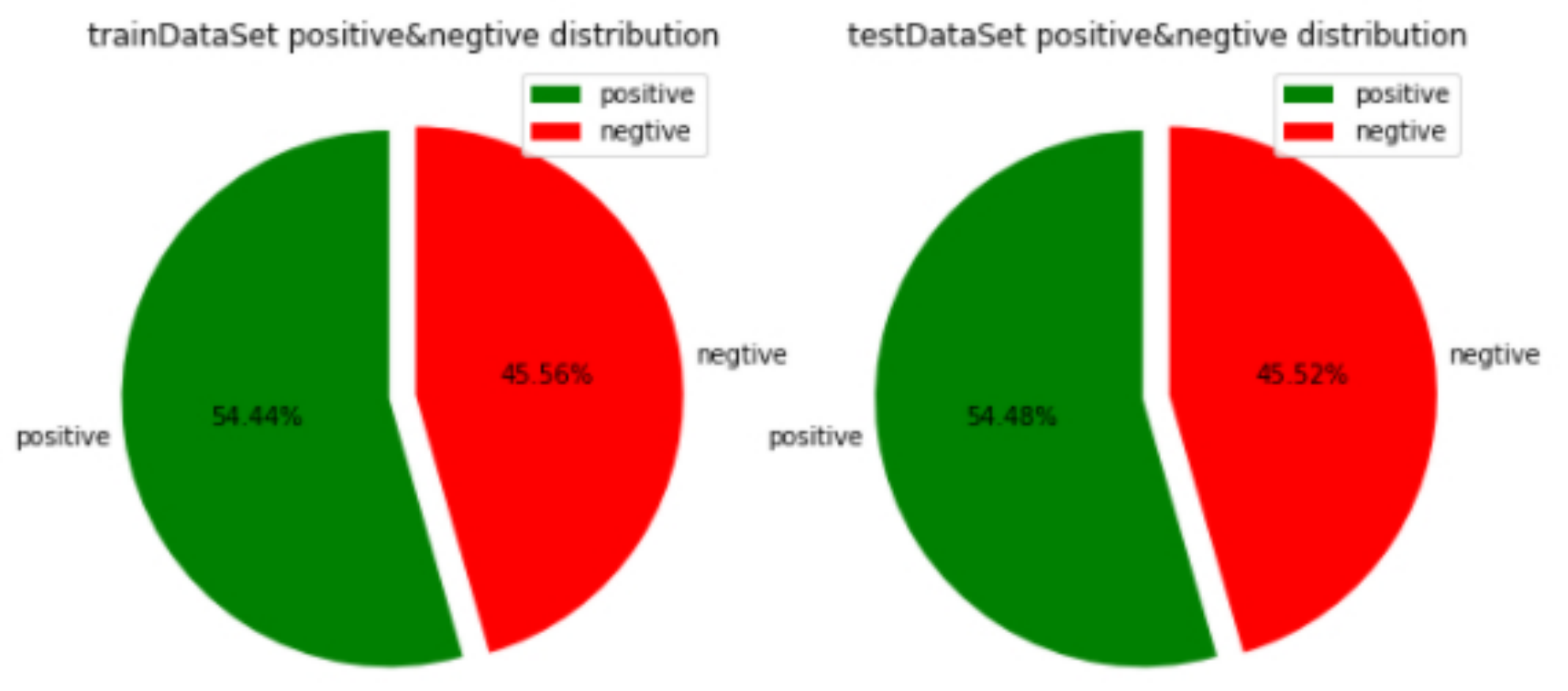}
	\end{center}
	\caption{Proportion of positive and negative samples of training dataset and test dataset}
	\label{fig:long}
	\label{fig:onecol}
\end{figure}

%-------------------------------------------------------------------------
\subsection{Baseline Algorithm}

In this experiment, we use the following cutting-edge deep learning recommendation algorithm as a baseline algorithm for experimentation:

• NeuralCF\cite{Authors19}: The simple dot product operation of the traditional matrix factorization algorithm is replaced with a multilayer neural network, which enables better cross-learning between user and item features, and enhances the model's fitting and expression ability.

• Embedding MLP\cite{Authors02}: Convert sparse one-hot features into dense embedding vectors and input them into a multilayer neural network to enhance the expressive ability of the model.

• DeepFM\cite{Authors21}: In the case of having prior knowledge, artificially combine and cross features instead of letting the neural network learn the cross combination by itself, which improves the accuracy and efficiency of model training.

• DIN\cite{Authors01}: Deep interest network, proposed by Alimama's algorithm team in KDD2018, uses humans' innate habit of attention, that is, increases attention weights, making the model more suitable for human beings.

%-------------------------------------------------------------------------
\subsection{Experiment Evirionments Setup and Ealuation Index}
This experiment runs in the following environment: operating system Ubuntu 18.04.5 LTS; Cuda 11.1; 8-core 16G memory; GeForce RTX 2060 super; 8G video memory; Python 3.8.1; Tensorflow 2.5; Spark 2.4.3.

In order to ensure the fairness of the experiment, this experiment and its baseline experiments all use batch size of 4500, Adam optimizer, and num\_buckets of users and movies of the same size. The experiment uses ROC AUC, PR AUC, F1-Score, Precision and Recall as the model effect evaluation indicators.

%-------------------------------------------------------------------------
\subsection{Evaluation of Experimental Results}
%-------------------------------------------------------------------------
\subsubsection{Comparison of Key Evaluation Indicators of Each Model}
This paper uses four cutting-edge deep learning recommendation algorithms such as  NeuralCF, Embedding MLP, DeepFM, DIN and so on as the baseline algorithm to compare with the MMDIN algorithm. Training and prediction are performed on the dataset described above, and finally, the evaluation indicators of the model are as follows in table 1:

\begin{table}[h]
	\renewcommand{\arraystretch}{1.3}
	\caption{Key evaluation indicators for each models}
	\label{tab1}
	\centering
	\scalebox{0.7}{  % resize table
	\begin{tabular}{cccccc}
		\hline
		\textbf{}             & \textbf{ROC-AUC} & \textbf{PR-AUC} & \textbf{F1} & \textbf{Precision} & \textbf{Recall} \\ \hline
		\textbf{NeuralCF}     & 0.7295           & 0.7516          & 0.7290      & 0.6717             & 0.7707          \\
		\textbf{EmbeddingMLP} & 0.7553           & 0.7731          & 0.7470      & 0.6880             & 0.7920          \\
		\textbf{DeepFM}       & 0.7802           & 0.7956          & 0.7540      & 0.7131             & 0.7881          \\
		\textbf{DIN}          & 0.7870           & 0.8011          & 0.7590      & 0.7184             & 0.7889          \\
		\textbf{MMDIN}        & 0.8006           & 0.8149          & 0.7670      & 0.7327             & 0.7883          \\ \hline
	\end{tabular}}
\end{table}

It can be seen from Table 1 that the NeuralCF model has the lowest ROC curve AUC value, PR AUC value, F1-Score, accuracy rate and recall rate. This reflects the shortcomings of the NeuralCF model --- NeuralCF interoperate the user and item side features for prediction, but the user and item side feature are static. Therefore,  it is difficult to integrate the scene features, so the indicators of NeuralCF are low; the advantage of NeuralCF is that the features obtained after training can be stored, and recommendation function can be realized with simple interoperability, and it is easy to infer and maintain online, so is also widely used. In the next is the Embedding MLP model, which reflects that the MLP model can fully cross different feature vectors, and has the characteristics of strong fitting and expression capabilities. Then is the DeepFM model, because the DeepFM model not only has the characteristics of the MLP model that can learn complex features and has strong fitting and expression capabilities but also can use the prior knowledge of people to artificially cross-combinate features to make the model effect more accurate and improves the efficiency of model training.

The DIN model has high values in all indicators, which reflects the improvement effect of the model after the attention mechanism is added to the model, which also makes the model more in line with the characteristics of human attention. On the MMDIN model, its recall rate and accuracy rate are about 0.7882 and 0.7327 respectively, both of which have the highest values. F1 value and PR-AUC value reflect the comprehensive effect of accuracy rate and recall rate. It is easy to know and MMDIN model also gets the highest value. ROC-AUC comprehensively reflects the model's prediction of false positive rate and true positive rate, and also has the highest value. These show that MMDIN has a better predictive effect than the previous four models, which also reflects the increasing effect of adding multi-head and multi-modal modules to the DIN model. In order to analyze the effect of the model more intuitively, 20,000 samples are sampled from the test dataset, and then analyze the PR curve, ROC curve and intuitive classification effect of the model.

%-------------------------------------------------------------------------
\subsubsection{Precision and Recall Rate Curve of Each Model}
The PR curve can intuitively indicate the change of accuracy rate with recall rate, and it is an important indicator to measure the effect of recommendation model. This experiment randomly sampled 20,000 pieces of data from the test dataset to predict and evaluate the model, and draw the PR curve as follows in figure 3:

\begin{figure}[h]
	\begin{center}
		\includegraphics[width=1.0\linewidth]{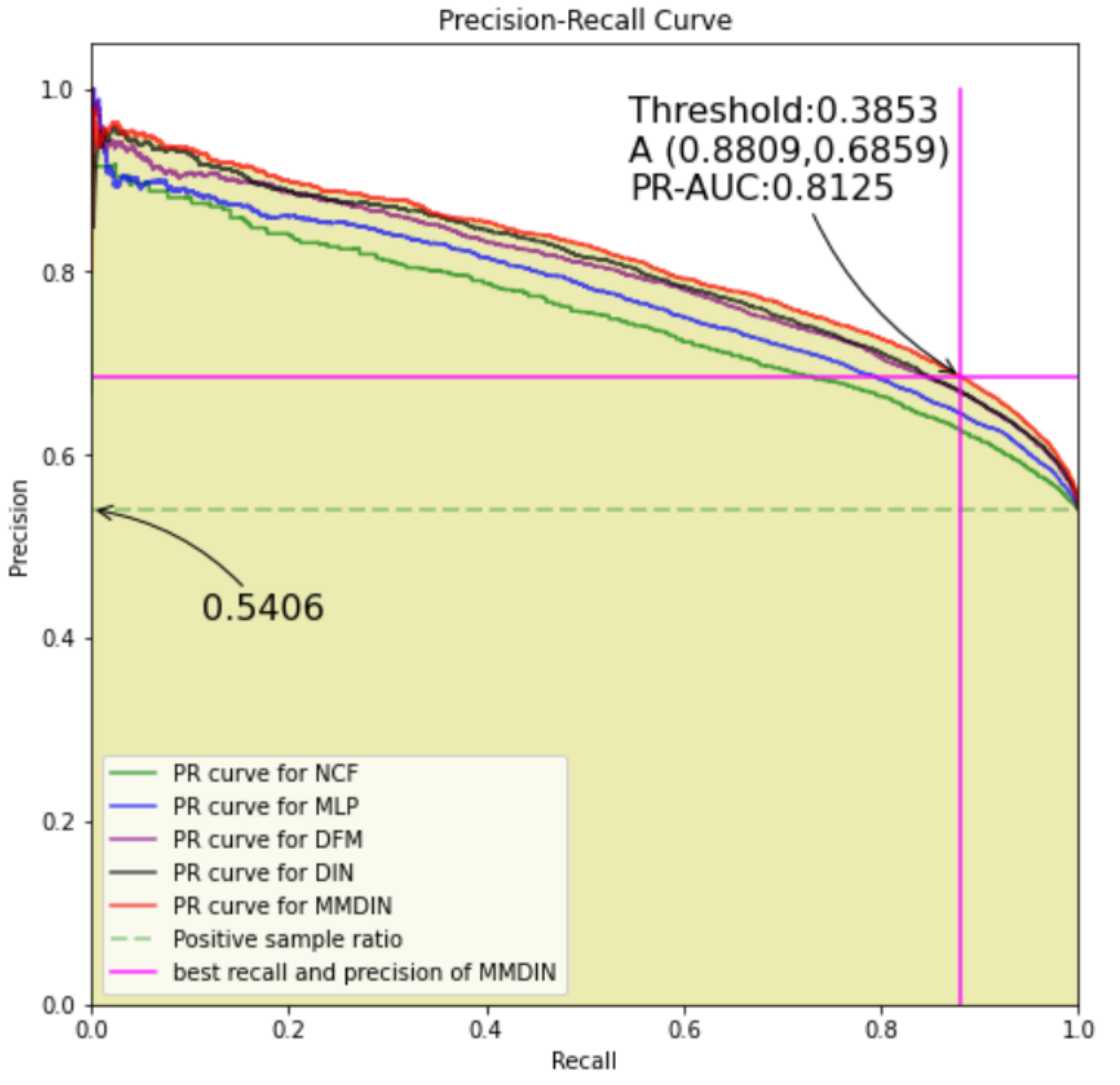}
	\end{center}
	\caption{Precision and Recall Rate Curve of Each Model}
	\label{fig:long}
	\label{fig:onecol}
\end{figure}

In figure 3, it can be seen from the green dotted line that the positive rate of the 20,000 samples is 0.5406, which is basically the same as the positive rate on the entire test dataset. Among them, the abscissa represents the recall rate, and the ordinate represents the accuracy rate. It can be seen from the figure that as the recall rate increases, the accuracy of each model continues to decrease. From bottom to top, the green curve at the bottom is NeuralCF, its overall effect is the worst; the blue, purple, and black curves is followed by, which are Embedding MLP, DeepFM, DIN, and they are basically the same as the evaluation results in the entire test dataset. The last is the red curve of MMDIN, which achieves the best comprehensive effect; it can be seen from the figure that at point A, that is, when the threshold is set to 0.3853, the model obtains the comprehensive optimal accuracy and recall rates, which are 0.6859 and 0.8809, respectively. The PR AUC value at this point is 0.8125, which is basically the same as the PR AUC value on the entire test dataset. As shown in figure 3, MMDIN has a better overall effect on recall and accuracy than other models.

%-------------------------------------------------------------------------
\subsubsection{Receiver Operating Curve of  Each Model}
The ROC curve is another important indicator for intuitive analysis to measure the quality of a recommendation model. It reflects the relationship between the false positive rate and the true positive rate, and is also called the "receiver characteristic working curve". The ROC curve drawn in this experiment for the evaluation of each model is as follows in figure 4:

\begin{figure}[h]
	\begin{center}
		\includegraphics[width=1.0\linewidth]{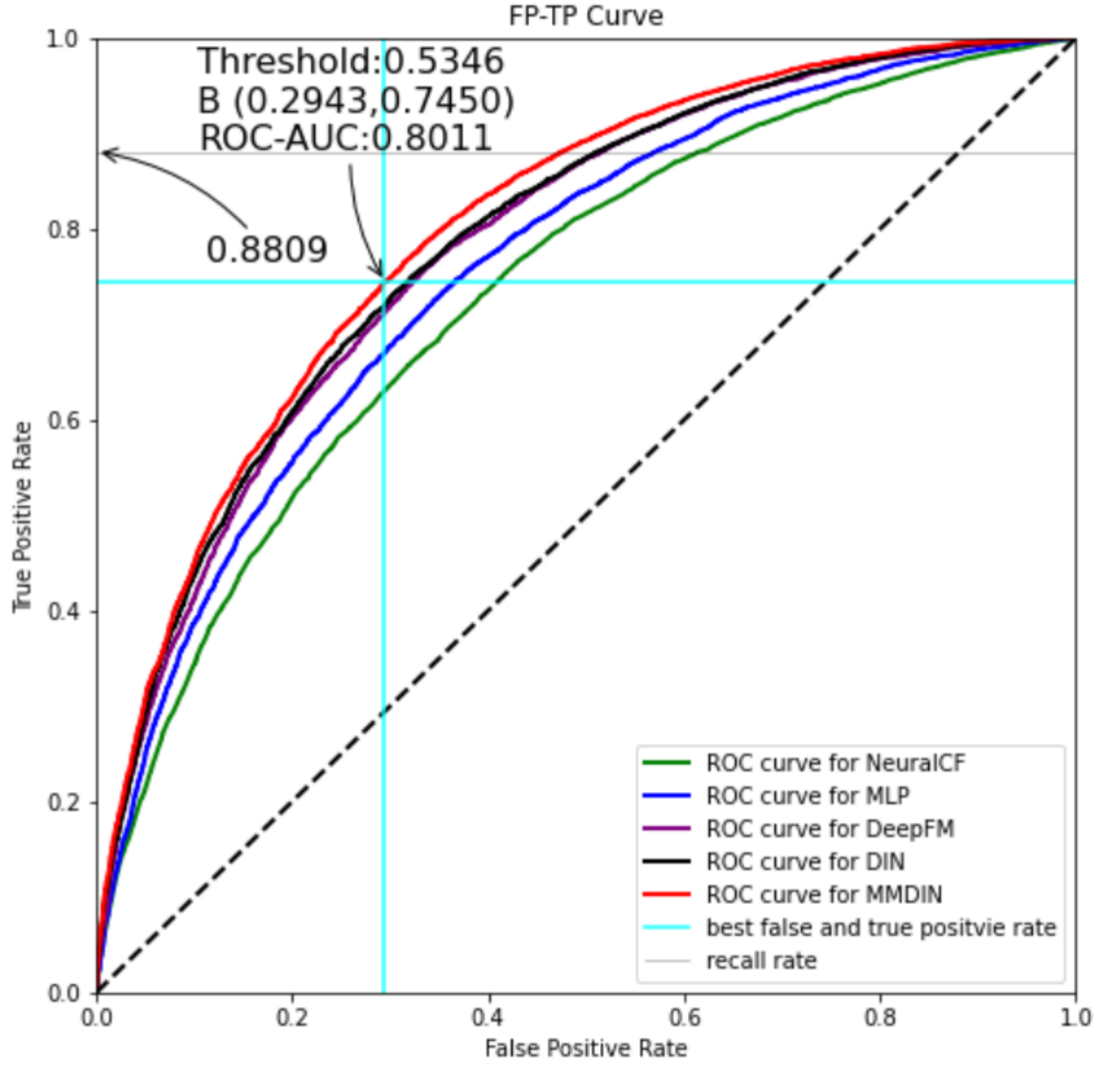}
	\end{center}
	\caption{Receiver Operating Curve of  Each Model}
	\label{fig:long}
	\label{fig:onecol}
\end{figure}

In figure 4, the abscissa is the false positive rate, and the ordinate is the true positive rate. It can be seen from the figure that as the false positive rate decreases, the true positive rate is also decreasing. From bottom to top, they are green, blue, purple, black, and red. They are the ROC curves of NeuralCF, MLP, DeepFM, DIN, and MMDIN. The overall effect is still from low to high: NeuralCF <Embedding MLP <DeepFM <DIN <MMDIN, which shows that the MMDIN model can achieve the best overall effect on false positive rate and true positive rate compared with other models. At the same time, it can be seen from the figure that at point B, that is, when the threshold value is 0.5346, the model obtains the optimal true positive rate and false positive rate, which are 0.7450 and 0.2943, respectively. At this time, the ROC AUC value is 0.8011, which is also basically the same as the ROC AUC values in the entire test dataset. The recall rate at this point is shown by the gray horizontal line in the figure, which is as high as 0.8809, which also achieves very good results.

%-------------------------------------------------------------------------
\subsubsection{Model Intuitive Classification Effect}
After analyzing the individual and comprehensive indicators above, let's take a look at the intuitive classification effect of the model. Randomly sample 20,000 pieces of data in the test dataset above, and use the MMDIN model to do classification prediction, and draw the results into a scatter plot as shown below:

\begin{figure}[h]
	\begin{center}
		\includegraphics[width=1.0\linewidth]{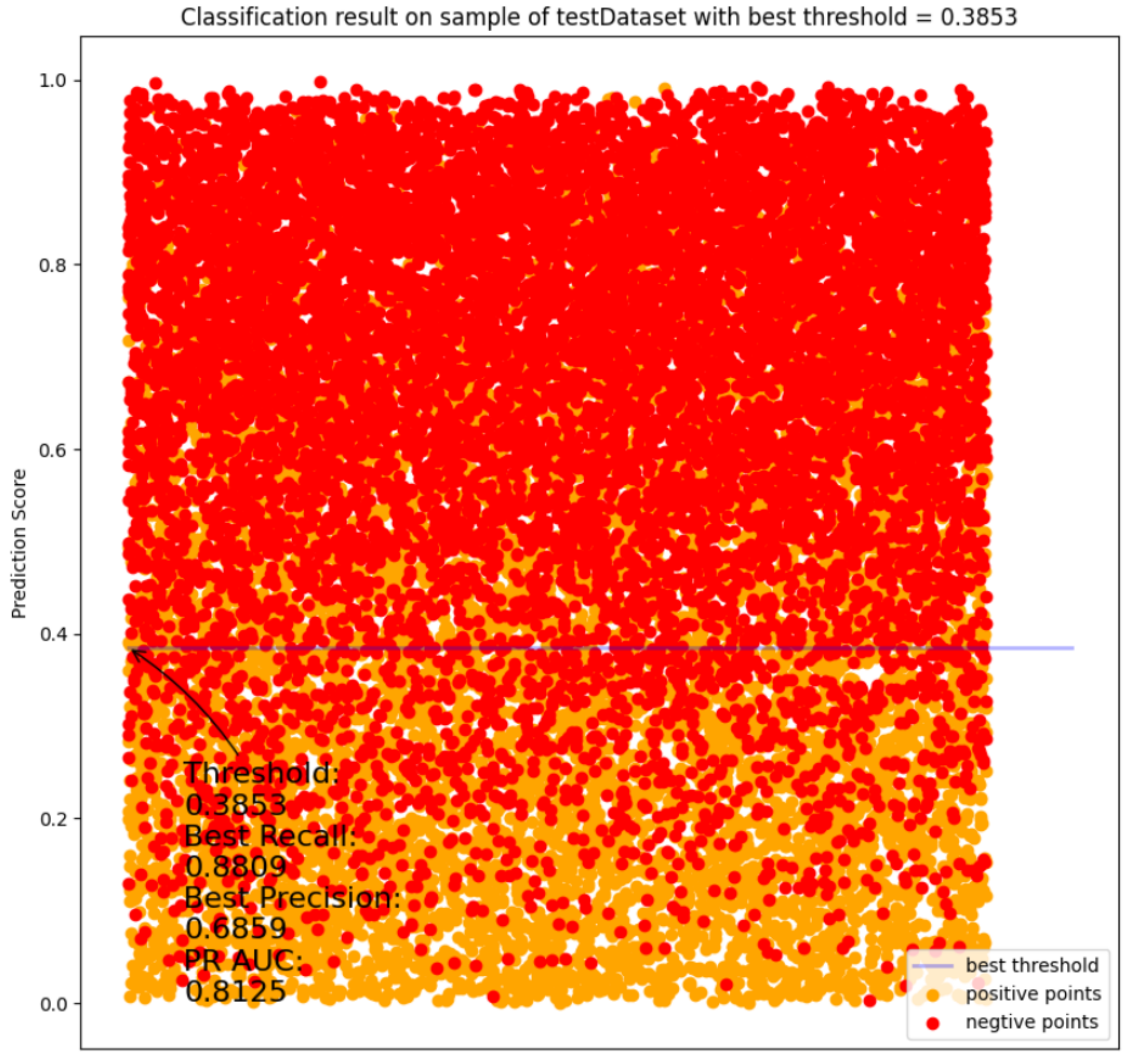}
	\end{center}
	\caption{Model intuitive classification effect}
	\label{fig:long}
	\label{fig:onecol}
\end{figure}

In figure 5, the ordinate indicates that the MMDIN model predicts the rating values of the user’s rating of the movie. The closer to 1, the higher the user’s rating of the movie; the abscissa is a random floating point number from 0 to 1, its purpose is to make the scatter not too much dense. The orange dot in the figure indicates that a user actually gave a negative review to a movie, and the red dot indicates that a user actually gave a good review to a movie. It can be intuitively felt from figure 5 that the points that are actually well received, the scores predicted by the model are also more larger, so the red points are denser in the upper area; in fact, the points that are bad reviews, the scores predicted by the model are also small, so the orange dots are denser in the lower area.

It can also be seen from figure 5 that when the threshold is 0.3853, the model obtains the optimal recall rate of 0.8809 and accuracy rate of 0.6859, and the PR AUC at this point is 0.8125. The division of the optimal threshold setting in figure 5 is also in line with intuitive perception, which shows that the prediction results of the MMDIN model are reliable.

%-------------------------------------------------------------------------
\subsection{The Effectiveness of MMDIN Model}
A detailed analysis of the PR and ROC curves has been done above, and the following is a comparison of the comprehensive indicators of each model, as shown in the figure below:

\begin{figure}[h]
	\begin{center}
		\includegraphics[width=1.0\linewidth]{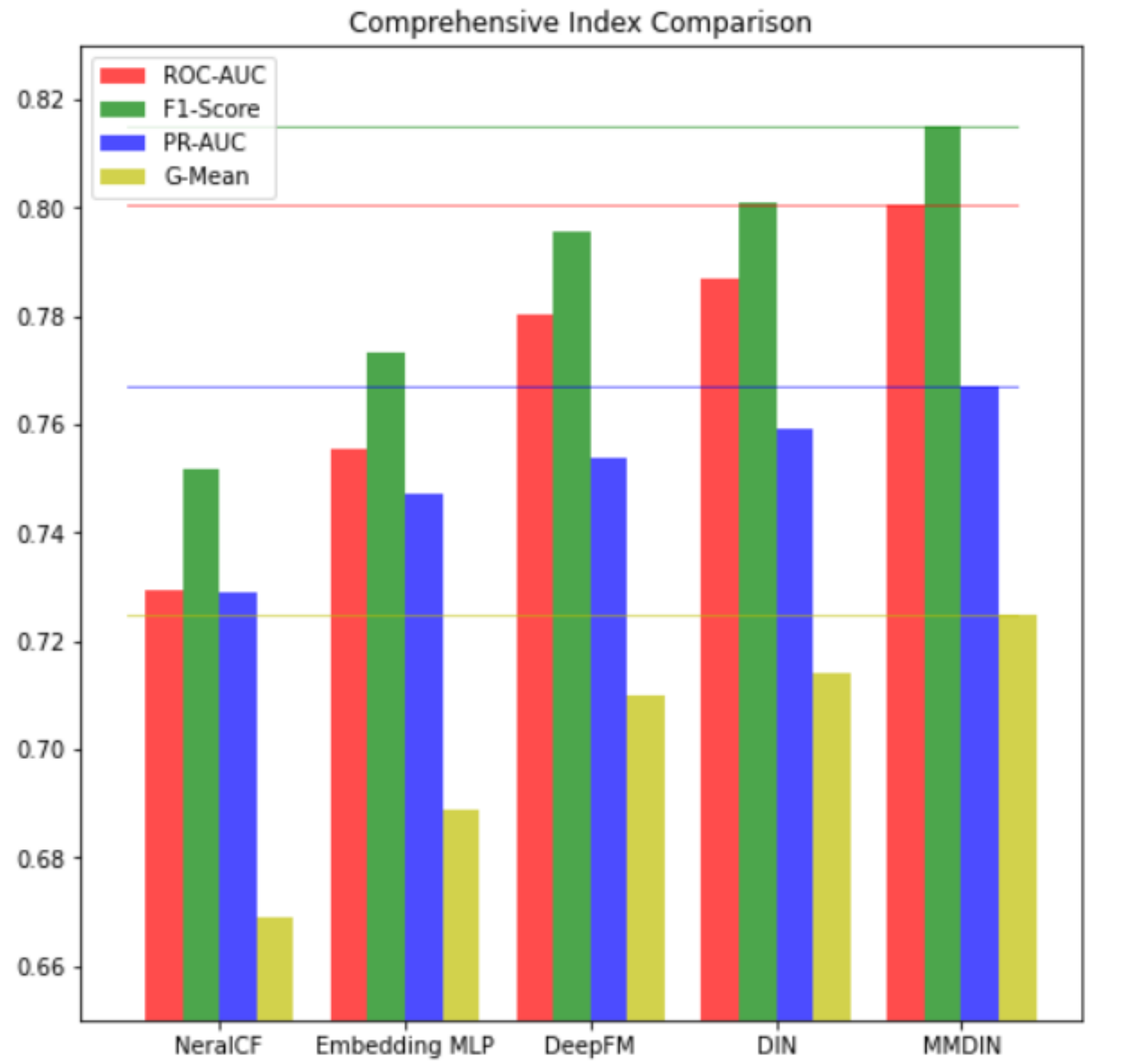}
	\end{center}
	\caption{Comparison of comprehensive indicators of each model}
	\label{fig:long}
	\label{fig:onecol}
\end{figure}

It can be seen from the figure above that the MMDIN model has the best effect than other models in all comprehensive indicators, which is mainly due to the addition of multi-head and multi-modal modules to the MMDIN model. The multi-modal module enriches the features that can be used for model training and prediction, making the final prediction result more accurate; and the multi-head mechanism enables the model to extract features and cross-combine data from different dimensions, making the model more robust in expressing. In summary, the comparison of key indicators of different models, PR curve, ROC curve, intuitive classification effect of the model, and comparison of comprehensive indicators shows that MMDIN has a better recommendation prediction effect than other models.

%-------------------------------------------------------------------------
\section{Conclusions}
In this article, we propose a new MMDIN model. The key to the MMDIN model is that a multi-head mechanism and a multi-modal module are added to the DIN model in contrast to models proposed in the past. The multi-head mechanism enables the model to extract features of the data from different dimensions, which improves the expressive ability of the model; while the multi-modal module extracts features from movie posters, which enriches the feature sets that the models can use to train and predict to improve the prediction effect. It can be seen from the above experiments that the MMDIN model has better predictive effects than other models on all comprehensive indicators.

{\small
\bibliographystyle{ieee.bst}
\bibliography{egpaper_final.bib}
}

\end{document}